\documentclass[intlimits,twoside,a4paper]{article}
\usepackage{amsmath,amssymb}
\usepackage{graphicx}
\usepackage[T2A]{fontenc}
\usepackage[cp1251]{inputenc}

\usepackage[eqsecnum]{cmpj2}

\DeclareMathOperator{\csch}{csch}

\issue{2014}{17}{1}{13703}
\doinumber{10.5488/CMP.17.13703}

\title[Complex conductivity in strongly fluctuating layered
superconductors]%
{Complex conductivity in strongly fluctuating layered
superconductors}
\author{B.D.~Tinh, L.M.~Thu, L.V.~Hoa}
\address{Department of Physics, Hanoi National University
of Education, 136 Xuanthuy, Caugiay, Hanoi, Vietnam}

\date{Received October 20, 2013, in final form February 7, 2014}
\begin{document}

\maketitle

\begin{abstract}
The time-dependent Ginzburg-Landau approach is used to calculate the complex
fluctuation conductivity in layered type-II superconductor under magnetic
field. Layered structure of the superconductor is
accounted for by means of the Lawrence-Doniach model, while the nonlinear
interaction term in dynamics is treated within self-consistent Gaussian
approximation.  In high-$T_{\mathrm{c}}$ materials, large portion of the $H-T$
diagram belongs to vortex liquid phase.  The expressions summing contributions of all the Landau levels are presented in explicit form which are applicable
essentially to the whole phase and are compared to experimental data on
high-$T_{\mathrm{c}}$ superconductor YBa$_{2}$Cu$_{3}$O$_{7-\delta }$. Above the crossover to the ``normal phase'',
our results agree with the previously obtained.
\keywords time-dependent Ginzburg-Landau, complex conductivity, type-II superconductor
\pacs 74.40.Gh, 74.25.fc, 74.20.De
\end{abstract}

\section{Introduction}

There has been a renewed interest in the effect of strong thermal fluctuations
 in layered high $T_{\mathrm{c}}$ superconductors that exhibit Nernst effect  \cite{Ong00} well above $T_{\mathrm{c}}$.
 The experiments were interpreted
microscopically as due to virtual (preformed) pairs  \cite{Varlamovmicro}, although can be described on the ``mesoscopic''
level using Ginzburg-Landau
approach  \cite{Huse}. This indicates large renormalization of $T_{\mathrm{c}}$ (of
order $1$) by thermal fluctuations in strongly layered materials like Bi$_{2}$Sr$_{2}$CaCuO$_{8+\delta }$ and La$_{2-x}$Sr$_{x}$CuO$_{4}$.
Microscopic parameters of the material determine both the actual $T_{\mathrm{c}}$ at which the Cooper pairs are coherent and the mean field
critical temperature $T_{\mathrm{c}}^{\mathrm{MF}}$ until which incoherent Cooper pairs exist
and effect the transport, magnetism and thermodynamics of the  material.
More recently strong diamagnetism was also observed  \cite{Ongmag}, although it
is still debated on experimental level. All these new
results are concerned with DC transport.

The AC transport in strongly fluctuating type-II
superconductors have been a subject for active research for many years, both
theoretically and experimentally. While the seminal calculation of the
enhancement of the DC conductivity in the normal phase due to virtual Cooper
pair created by thermal fluctuations by Aslamazov and Larkin was done in the
framework of the microscopic BCS theory  \cite{Varlamov}, the approach
becomes cumbersome in more complicated situations involving external
magnetic field, layerred structure, etc. As usual in these circumstances
(especially in the absense of a simple accepted microscopic model for high $%
T_{\mathrm{c}}$ and other recently discovered ``unconventional'' strongly fluctuating
superconductors), a more phenomenological Ginzburg-Landau approach adapted
to incorporate thermal fluctuations turns out to be more effective  \cite{Varlamov,Rosenstein10}.
A general method to model the thermal fluctuations in dynamics is to add a random Langevin white
noise  to the time-dependent Ginzburg-Landau (TDGL) equations. The transport coefficients within this approach are obtained as a
long time limit of a driven system. The model, therefore, becomes rather
complicated and approximations should be made. In the gaussian fluctuations
regime (in the normal phase not very close to criticality, where the quartic
term in the GL free energy is dominant), the expressions for complex
conductivity at zero magnetic field have been obtained very early on  \cite%
{Schmidt,Schmidt2}. This was expanded later by Dorsey and coworkers  \cite{Dorsey91,Wick00} to include the critical fluctuations region by a variety of
nonperturbative methods (Hartree approximation, large number of components $%
N $ limit, $\varepsilon $-expansion. The results were in line  \cite{Dorsey91} with general physical scaling arguments by Fisher, Fisher, and
Huse  \cite{Fisher91}.

The complex conductivity in magnetic field in the normal phase was calculated using TDGL equation by Larkin and Varlamov
 \cite{Varlamov}. The general expression valid for complex conductivity in layered
superconductors under the assumption of gaussian fluctuations (neglecting the
quartic in the order parameter term in the GL free energy) was also calculated
and presented in  \cite{Varlamov} as a sum over all the Landau levels. On the
opposite side of the phase diagram, namely in a strongly pinned case (vortex
glass and Bragg glass), the same quantity was calculated using both
macroscopic elastic theory  \cite{Ong97} and the TGDL  \cite{Maniv}. In yet
another limit of the Abrikosov lattice phase of a clean superconductor, see  \cite{Rosenstein10}, the complex conductivity was recently calculated  \cite{Lin09}
(within the lowest Landau level). The present work is complementary to all these
in that we concentrate on the thermally depinned homogeneous phase marked as
``vortex liquid''  \cite{Rosenstein10}. The complex conductivity in magnetic
field in 2D and 3D was calculated using TDGL equation  \cite{Tinh12,Tinh13}. In
strongly layered high $T_{\mathrm{c}}$ materials, this portion of the magnetic phase
diagram is very large and consequently well studied experimentally  \cite{Yoshi01,Hanaguri99,Brbic}. In this region, thermal fluctuations are so strong that
one cannot neglect the quartic in the order parameter term of the GL energy.
This term, however, can be incorporated self-consistently into the framework
of  \cite{Varlamov}.

In this paper, the complex conductivity including all Landau
levels is calculated in a layered superconductor under magnetic field in the vortex liquid
phase by using TDGL approach with thermal fluctuations modelled by the
Langevin white noise. We obtain an expression summing all Landau levels in an explicit form. The rest of the paper is
organized as follows. In section~2, the Lawrence-Donich model in its time
dependent form is briefly recalled and the main assumptions are specified. In
section~3, the interaction term in dynamics is treated within
self-consistent gaussian approximation sufficient for description of the
vortex liquis. The complex conductivity calculation within the same approximation
is the subject of section~4. The results are compared with experimental
data on HTSC in section~5, while the work is summarized in section~6.

\section{Thermal fluctuations in the time dependent GL Lawrence-Doniach model}
Cooper pairing in layered superconductors can be described by the 2D
distribution of the order parameter $\Psi _{n}\left( \mathbf{r}\right) $ in
each of the layers labeled by $n$. The Lawrence-Doniach version of the GL
free energy includes the Josephson coupling between the layers  \cite{Varlamov}:
\begin{equation}
F_{\mathrm{GL}} =s'\sum_{n}\int \rd^{2}r\left(\frac{{\hbar }^{2}}{2m^{\ast }}%
|\mathbf{D}\Psi _{n}|^{2}+\frac{{\hbar }^{2}}{2m_{\mathrm{c}}d'^{2}}|\Psi
_{n}-\Psi _{n+1}|^{2}+a|\Psi _{n}|^{2}+\frac{b'}{2}|\Psi _{n}|^{4}\right).
\label{original.Fre}
\end{equation}%
Here, $s'$ is the order parameter effective \textquotedblleft
thickness'' and $d'>s'$ is the distance between
layers. The Lawrence-Doniach model approximates the paired electrons density
of states by homogeneous infinitely thin planes separated by distance $%
d'$. For simplicity, we assume $a=\alpha T_{\mathrm{c}}^{\mathrm{MF}}(t-1)\,$, $t^{\mathrm{MF}}\equiv T/T_{\mathrm{c}}^{\mathrm{MF}}$, although this temperature dependence can be
easily modified to better describe the experimental coherence length. The
``mean field'' critical temperature $%
T_{\mathrm{c}}^{\mathrm{MF}}$ depends on UV cutoff and is often much larger than
``renormalized'' critical temperature $T_{\mathrm{c}}$. This temperature is
significantly higher than the measured critical temperature $T_{\mathrm{c}}$ due to
strong thermal fluctuations on the mesoscopic scale.

The covariant derivatives are defined by $\mathbf{D}\equiv \pmb{\nabla }%
+\ri(2\pi /\Phi _{0})\mathbf{A},$ where the vector potential describes a
constant and homogeneous magnetic field $\mathbf{A}=\mathbf{(-}By,0\mathbf{)}
$ and $\Phi _{0}=hc/e^{\ast }$ is the flux quantum with $e^{\ast
}=2\left\vert e\right\vert $. The two scales, the coherence length $\xi ^{2}=%
{\hbar }^{2}/(2m^{\ast }\alpha T_{\mathrm{c}}),$ \ and the penetration depth,\ \ $%
\lambda ^{2}=c^{2}m^{\ast }b'/(4\pi e^{\ast 2}\alpha T_{\mathrm{c}})$ define
the GL ratio $\kappa \equiv \lambda /\xi $, which is very large for HTSC. In
the case of strongly type-II superconductors, the magnetization is by a
factor $\kappa ^{2}$ smaller than the external field for magnetic field
larger than the first critical field $H_{\mathrm{c}1}\left( T\right) $, so that we
take $B\approx H$. The electric current, $\mathbf{J}=\mathbf{J}^{\mathrm{n}}+\mathbf{J%
}^{\mathrm{s}}$, includes both the Ohmic normal part
\begin{equation}
\mathbf{J}^{\mathrm{n}}=\sigma _{n}\mathbf{E}\,,  \label{Ohm}
\end{equation}%
and the supercurrent
\begin{equation}
\mathbf{J}_{n}^{\mathrm{s}}\left( \mathbf{r}\right) =\frac{\ri e^{\ast }{\hbar }}{%
2m^{\ast }}\left( \Psi _{n}^{\ast }\mathbf{D}\Psi _{n}-\Psi _{n}\mathbf{D}%
\Psi _{n}^{\ast }\right) .  \label{Js_I}
\end{equation}%
Since we are interested in a transport phenomenon, it is necessary to
introduce a dynamics of the order parameter. The simplest one is a
gauge-invariant version of the ``type A''
relaxational dynamics  \cite{Ketterson}. In the presence of thermal fluctuations, which on the
mesoscopic scale are represented by a complex white noise, it reads:
\begin{equation}
\frac{{\hbar }^{2}\gamma '}{2m^{\ast }}D_{\tau }\Psi _{n}=-\frac{1}{%
s'}\frac{\delta F_{\mathrm{GL}}}{\delta \Psi _{n}^{\ast }}+\zeta _{n}\,,
\label{TDGL_I}
\end{equation}%
with correlator
\begin{equation}
\left\langle \zeta _{n}\left( \mathbf{r,}\tau \right)
,\zeta _{m}\left( \mathbf{r}'\mathbf{,}\tau '\right)
\right\rangle =\frac{{\hbar }^{2}\gamma 'T}{m^{\ast }s'}\delta _{nm}\delta \left( \mathbf{r-r}'\right) \delta
\left( \tau -\tau '\right) .
\end{equation}
 Here, $D_{\tau }\equiv \partial/\partial \tau -\ri(e^{\ast }/\hbar )\Phi $ is the covariant time derivative,
with $\Phi =-E_{\tau }y\,$ being the scalar electric potential describing the
driving force in a purely dissipative dynamics. The electric field is, to a
good approximation in the vortex liquid phase coordinate, independent (at
least for frequences below THz range, see argumentation in  \cite%
{Rosenstein07}), but is a monorchromatic periodic function of time%
\begin{equation}
E_{x}=0, \qquad E_{y}(\tau )=E\exp (-\ri\omega \tau ).  \label{E}
\end{equation}

Throughout most of the paper we use the coherence length $\xi $ as a unit of
length and $H_{\mathrm{c}2}=\Phi _{0}/2\pi \xi ^{2}$ as a unit of the magnetic field,
with dimensionles field $b=B/H_{\mathrm{c}2}$. In analogy to the coherence length, one
defines a characteristic time scale: the GL \textquotedblleft
relaxation'' time $\tau _{\mathrm{GL}}=\gamma '\xi ^{2}/2$.
Similarly, it is convenient to use the following unit of the electric field,
$E_{\mathrm{GL}}=H_{\mathrm{c}2}\xi /c\tau _{\mathrm{GL}}\,$, so that the dimensionless field is
$\mathcal{E}=E/E_{\mathrm{GL}}$. The dynamical equation, equation~(\ref{TDGL_I}), written in
dimensionless units reads:

\begin{equation}
\left( D_{\tau }-\frac{1}{2}D^{2}\right) \psi _{n}+\frac{1}{2d^{2}}(2\psi
_{n}-\psi _{n+1}-\psi _{n-1}) -\frac{1-t^{\mathrm{MF}}}{2}\psi _{n}+|\psi
_{n}|^{2}\psi _{n}=\overline{\zeta }_{n}\,,  \label{TDGL2lac}
\end{equation}
where $d=d'/\xi _{z}$ is dimensionless layer distance. The
coherence length perpendicular to the layers is smaller compared to $\xi $
by the anisotropy parameter $\gamma $ \cite{Rosenstein10}.

The covariant time derivatives become $D_{\tau }={\partial }/{\partial
\tau }+\ri\mathcal{E}(\tau )y$, the covariant derivatives are defined by $D_{x}={\partial }/{\partial x}-\ri by\,$, $D_{y}={\partial }/{\partial y}$. The ``mean field'' critical temperature $T_{\mathrm{c}}^{\mathrm{MF}}$ depends on the ultraviolet (UV) cutoff. This temperature is higher
than the measured critical temperature $T_{\mathrm{c}}$ due to strong thermal
fluctuations on the mesoscopic scale, and it will be renormalized later. The
dimensionless Langevin white-noise forces, $\overline{\zeta }_{n}=b'^{1/2}\left(2\alpha T_{\mathrm{c}}^{\mathrm{MF}}\right)^{-3/2}\zeta _{n}$, are correlated through $%
\left\langle \overline{\zeta }_{n}^{\ast }(\mathbf{r},\tau )\overline{\zeta }%
_{m}(\mathbf{r}',\tau ')\right\rangle =2\eta t/s\delta
_{nm}\delta (\mathbf{r}-\mathbf{r}')\delta (\tau -\tau ')$, with a dimensionless fluctuation strength parameter related to the well
known Ginzburg number  \cite{Varlamov,Rosenstein10} by
\begin{equation}
\eta =\pi \sqrt{2\mathrm{Gi}}\,, \qquad \mathrm{Gi}=\frac{1}{2}\left( \frac{8e^{2}\kappa
^{2}\xi T_{\mathrm{c}}^{\mathrm{MF}}\gamma }{c^{2}{\hbar }^{2}}\right) ^{2}.  \label{eta}
\end{equation}

The dimensionless current density is $\mathbf{J}^{\mathrm{s}}=J_{\mathrm{GL}}\mathbf{j}^{\mathrm{s}}$,
where
\begin{equation}
\mathbf{j}_{n}^{\mathrm{s}}=\frac{\ri}{2}\left( \psi _{n}^{\ast }%
\mathbf{D}\psi _{n}-\psi _{n}\mathbf{D}\psi _{n}^{\ast }\right)
\label{current}
\end{equation}%
with $J_{\mathrm{GL}}=cH_{\mathrm{c}2}/(2\pi \xi \kappa ^{2})$ being the unit of the current
density. Consistently, the conductivity will be given in the units of
\begin{equation}
\sigma _{\mathrm{GL}}=\frac{J_{\mathrm{GL}}}{E_{\mathrm{GL}}}=\frac{c^{2}\gamma '}{4\pi \kappa ^{2}}\,.
\label{unitcond}
\end{equation}%
This unit is close to the normal state conductivity $\sigma _{n}$ in dirty
limit superconductors  \cite{Kopnin}. In general, there is a factor $k$ of
the order one relating the two: $\sigma _{n}=k\sigma _{\mathrm{GL}}$.
\section{The Green's function of TDGL in Gaussian approximation}

Let us first assume that the vortex liquid is not driven by the electric
field. As mentioned above, the cubic term in the TDGL equation~(\ref{TDGL2lac})
can be treated in the self-consistent gaussian approximation (explained in
detail in  \cite{Tinh10}) by replacing $|\psi _{n}|^{2}\psi _{n}$ with a
linear one $2\left\langle |\psi _{n}|^{2}\right\rangle \psi _{n}$
\begin{equation}
\left( \frac{\partial }{\partial \tau }-\frac{1}{2}D^{2}-\frac{b}{2}\right)
\psi _{n}+\frac{1}{2d^{2}}(2\psi _{n}-\psi _{n+1}-\psi _{n-1})+\varepsilon
\psi _{n}=\overline{\zeta }_{n}\,.  \label{TDGL2lacr}
\end{equation}%
Here, the value of the coefficient of the linear term,
\begin{equation}
\varepsilon =-\frac{1-t^{\mathrm{MF}}-b}{2}+2\left\langle |\psi _{n}|^{2}\right\rangle,  \label{gap.eq1}
\end{equation}%
is different from the noninteracting one.

The relaxational linearized TDGL equation with a Langevin noise, equation~(\ref{TDGL2lacr}), is solved using the retarded (vanishing for $\tau <\tau
'$) Green function (GF) $G_{k_{z}}^{0}(\mathbf{r},\tau ;\mathbf{r}%
',\tau ')$:
\begin{equation}
\psi _{n}(\mathbf{r},\tau ) =\int_{0}^{2\pi /d}\frac{\rd k_{z}}{2\pi }%
\re^{-\ri nk_{z}d}\int \rd \mathbf{r}' \int \rd\tau 'G_{k_{z}}^{0}\left(\mathbf{r},\tau ;\mathbf{r}%
',\tau '\right)\overline{\zeta }_{k_{z}}\left(\mathbf{r}',\tau '\right).
\end{equation}%
The GF satisfies
\begin{equation}
\left\{ \frac{\partial }{\partial \tau }-\frac{1}{2}D^{2}-\frac{b}{2}+%
\frac{1}{d^{2}}[1-\cos (k_{z}d)]+\varepsilon \right\} G_{k_{z}}^{0}(\mathbf{r},\mathbf{r}',\tau -\tau')=\delta (\mathbf{r}-\mathbf{r}')\delta (\tau -\tau ').
\label{GFdef}
\end{equation}%
The GF is a Gaussian
\begin{equation}
G_{k_{z}}^{0}\left( \mathbf{r},\mathbf{r}',\overline{\tau}\right) =\theta \left( \overline{\tau}\right)C_{k_{z}}\exp \left[ \frac{\ri b}{2}X\left( y+y'\right) \right]  \exp \left( -\frac{%
X^{2}+Y^{2}}{2\beta }\right)
 ,  \label{Ansatz}
\end{equation}%
where $X=x-x',Y=y-y', \overline{\tau}=\tau -\tau '.$ $\theta \left( \overline{\tau}\right) $ is the Heaviside step function, $C$ and $\beta $ are coefficients.

Substituting equation~(\ref{Ansatz}) into equation~(\ref{GFdef}), one obtains:
\begin{equation}
\beta =\frac{2}{b}\tanh \left( b \overline{\tau}/2\right) ,
\label{beta.fun}
\end{equation}%
\begin{equation}
C =\frac{b}{4\pi }\exp \left( -\left\{ \varepsilon -\frac{b}{2}+\frac{1}{d^{2}}\left[1-\cos(k_{z}d)\right]\right\} \overline{\tau}\right)
\left[\sinh\left( \frac{b\overline{\tau}}{2}\right)\right]^{-1} .
\label{C.fun}
\end{equation}%

The thermal average of the density of Cooper pairs can
be expressed via the Green's functions:
\begin{eqnarray}
\left\langle \left\vert \psi _{n}(\mathbf{r},\tau )\right\vert
^{2}\right\rangle &=&2\omega t\frac{d}{s}\int_{0}^{2\pi /d}\frac{\rd k_{z}}{2\pi } \int \rd \mathbf{r}' \int \rd\tau '\left\vert G_{k_{z}}^{0}(\mathbf{r-r}',\tau -\tau ')\right\vert ^{2}  \nonumber \\
&=&\frac{\omega tb}{2\pi s}\int_{\overline{\tau} =\tau _{\mathrm{c}}}^{\infty }\frac{%
f(\varepsilon ,\overline{\tau} )}{\sinh (b\overline{\tau} )}\,,  \label{expect.v}
\end{eqnarray}%
where
\begin{equation}
f(\varepsilon ,\overline{\tau} ) =\exp \left[ \frac{2v^{2}}{b}\tanh \left( \frac{%
b\overline{\tau} }{2}\right) \right] \re^{-\left( 2\varepsilon -b+v^{2}\right) \overline{\tau} }
\re^{-2\overline{\tau} /d^{2}}I_{0}\left( 2\overline{\tau} /d^{2}\right).  \label{f}
\end{equation}%

Substituting the density of Cooper pairs (\ref{expect.v}) into equation~(\ref{gap.eq1}) and after renormalization  \cite{Tinh10},
the latter takes the form
\begin{eqnarray}
\varepsilon =-\frac{1-b-t}{2}-\frac{\eta t}{\pi s}\int_{0}^{\infty }\rd \overline{\tau }\,\ln [\sinh (b\overline{\tau })]\frac{\rd}{\rd\overline{\tau }}%
\left[ \frac{g(\varepsilon ,\overline{\tau })}{\cosh (b\overline{\tau })}%
\right]+\frac{\omega t}{\pi s}\left[ \gamma _{\mathrm{E}}-\ln (bd^{2})\right] ,\nonumber \\
\label{gapequation}
\end{eqnarray}%
where $g(\varepsilon ,\overline{\tau })=\re^{-\left( 2\varepsilon
-b+2/d^{2}\right) \overline{\tau }}I_{0}\left( {2\overline{\tau }}{d^{-2}%
}\right) $ with $I_{0}(x)=(1/2\pi )\int_{0}^{2\pi }\re^{x\cos \theta }\rd\theta$ being the modified Bessel function, $\overline{\tau }=\tau -\tau '$%
, $t=T/T_{\mathrm{c}}$, and $\gamma _{\mathrm{E}}=0.577$ is Euler constant. As was discussed
in  \cite{Tinh10} there are UV divergences in the intermediate steps, equation~(\ref{gap.eq1}), that are regularized by a cutoff. There is a degree of
arbitrariness in its choice and it was shown that in the TDGL theory, the
cutoff having a dimension of time, $\tau _{\mathrm{c}}$, rather than energy, is most
convenient. However, in the renormalized equation, equation~(\ref{gapequation}),
the cutoff does not appear: the mean field temperature $T_{\mathrm{c}}^{\mathrm{MF}}$ is now
replaced by $T_{\mathrm{c}}$. Having determined the GF, one can use it to calculate
the complex conductivity.

\section{The complex conductivity of a layered superconductor}

After the thermal averaging of the supercurrent density at time $\tau $, defined
by equation~(\ref{current}), it can be expressed via the Green's functions as follows:
\begin{equation}
j_{y}^{\mathrm{s}}\left( \tau \right) =\ri\eta t\frac{d}{s}\int_{0}^{2\pi /d}\frac{%
\rd k_{z}}{2\pi }\int_{r',\tau '}G_{k_{z}}^{\ast }\left(
\mathbf{r},\mathbf{r}',\tau -\tau '\right)  \frac{\partial }{\partial y}G_{k_{z}}\left( \mathbf{r},\mathbf{r}%
',\tau -\tau '\right) +\textrm{c.c.},  \label{jy_ac}
\end{equation}%
where $G_{k_{z}}\left( \mathbf{r},\mathbf{r}',\tau -\tau '\right) $ is the Green's function of the linearized TDGL equation~(\ref{TDGL2lac}%
) in the presence of the scalar potential describing the electric field. The
correction to the Green's function to linear order in the time dependent
homogeneous electric field $\mathcal{E}(\tau )$ is as follows:
\begin{equation}
G_{k_{z}}(\mathbf{r},\mathbf{r}',\tau -\tau ' ) =G_{k_{z}}^{0}(\mathbf{r},%
\mathbf{r}',\tau -\tau ' )-\ri\int \rd\mathbf{r}_{1}\int \rd\tau _{1}
\label{full_Green}
 G_{k_{z}}^{0}(\mathbf{r},\mathbf{r}_{1},\tau -\tau _{1}) \mathcal{E}%
(\tau _{1})y_{1}G_{k_{z}}^{0}(\mathbf{r}_{1},\mathbf{r}',\tau
_{1} -\tau ').
\end{equation}
Substituting the full Green function (\ref{full_Green}) into expression (\ref%
{jy_ac}), and performing the integrals, one obtains:
\begin{eqnarray}
j_{y}^{\mathrm{s}}(\tau ) &=&\frac{b}{4\pi s}\frac{\eta t}{b^{2}+\omega ^{2}}\mathcal{%
E}\int_{0}^{\infty }\rd\overline{\tau }\exp \left[ -\left( 2\varepsilon -b+\frac{2}{d^{2}}\right)
\overline{\tau }\right]I_{0}\left( \frac{2%
\overline{\tau }}{d^{2}}\right)\csch(b\overline{\tau }) \nonumber \\ &&
\times \left\{b\cos(\tau \omega)\cos(b\overline{\tau })-b\cos
[ (\tau-\overline{\tau })\omega]\csch(b\overline{\tau })\right.+\left.\omega \sin (\tau \omega) \right\} .
\label{finalcurrent}
\end{eqnarray}%
The real and imaginary parts of the complex conductivity after the Fourier transform
\begin{equation}
\sigma _{\mathrm{s}}(\omega )=\frac{j_{\mathrm{s}}(\omega )}{\mathcal{E(\omega )}}=\sigma
_{1}(\omega )+\ri\sigma _{2}(\omega ),  \label{complex_conduc}
\end{equation}%
therefore, are as follows:
\begin{equation}
\sigma _{1}(\omega ) =\frac{b^{2}}{8\pi s}\frac{\eta t}{b^{2}+\omega ^{2}}%
\int_{0}^{\infty }\rd\overline{\tau }\exp \left[ -\left( 2\varepsilon -b+\frac{2}{d^{2}}\right)
\overline{\tau }\right]I_{0}\left( \frac{2\overline{\tau }}{d^{2}%
}\right) \csch ^{2}(b\overline{\tau })
 [\cosh (b\overline{\tau })-\cos (\omega \overline{\tau })]
 ,  \label{real}
\end{equation}%
\begin{equation}
\sigma _{2}(\omega ) =\frac{b}{8\pi s}\frac{\eta t}{b^{2}+\omega ^{2}}%
\int_{0}^{\infty }\rd\overline{\tau }\exp \left[ -\left( 2\varepsilon -b+\frac{2}{d^{2}}\right)
\overline{\tau }\right]I_{0}\left( \frac{2\overline{\tau }}{d^{2}%
}\right)\csch ^{2}(b\overline{\tau })
 [\omega \sinh (b\overline{\tau })-b\sin (\omega \overline{\tau })] .  \label{imaginary}
\end{equation}
This is the main result of the paper. The complex resistivity of the layered
material is (after noting that superconducting layers constitute a fraction $%
s/d$ of the material)%
\begin{equation}
\sigma \left( \omega \right) =\frac{s}{d}\sigma _{\mathrm{s}}\left( \omega \right)
+\sigma _{n}\left( \omega \right).
\end{equation}
Neglecting the normal part, one obtains resistivity:%
\begin{equation}
\rho _{\mathrm{s}}(\omega )=\frac{d}{s\sigma _{\mathrm{s}}(\omega )}=\rho _{1}-\ri\rho _{2}\,,
\end{equation}%
where
\begin{equation}
\rho _{1}=\frac{d}{s}\frac{\sigma _{1}}{\sigma _{1}^{2}+\sigma _{2}^{2}}\,, \qquad \rho _{2}=\frac{d}{s}\frac{\sigma _{2}}{\sigma _{1}^{2}+\sigma
_{2}^{2}}\,.
\end{equation}%
The difference between the conductivities in a zero field and in the field of $B$ is defined as follows:
\begin{equation}
\mathbf{\delta }\sigma=\sigma(0)-\sigma(B).
\label{conduc}
\end{equation}

\section{Comparison with experiment}
\begin{figure}
\begin{center}
\includegraphics[angle = 0, width = 0.55\textwidth]{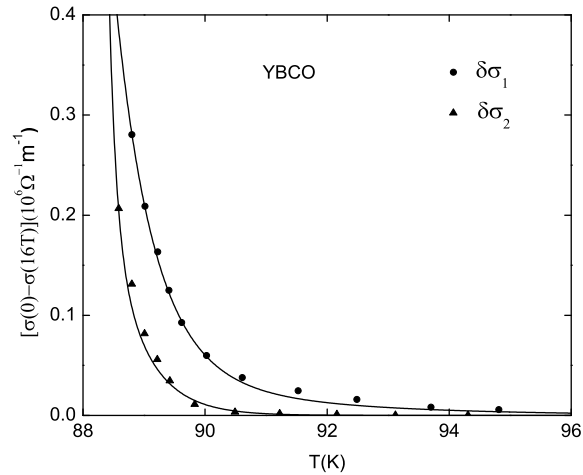}
\end{center}
\caption{{\protect\small Points are the difference between the conductivities in zero field and in the field of
16 T ($\mathbf{\delta }\sigma_{1}$ black circles, $\mathbf{\delta }\sigma_{2}$ black triangles) of slightly underdoped
YBCO. The solid lines are the theoretical values of resistivity for different temperatures at frequency $\omega
/2\pi =15.15$~GHz calculated from equation~(\protect\ref{conduc}) with
fitting parameters (see text).}}
\label{ac}
\end{figure}
The experimental results by M.S.~Grbi\'{c} et al. \cite{Brbic},
obtained from the the microwave absorption measurements at $\omega
/2\pi =15.15$~GHz on slightly underdoped YBa$_{2}$Cu$_{3}$O$_{7-\delta }$ (YBCO) with $T_{\mathrm{c}}=87$~K. The distance between the bilayers using the calculation is $d'=11.68$~\AA{} in  \cite{Yan91}. In order to compare the fluctuation conductivity with
experimental data in HTSC, one cannot use the expression of relaxation time $%
\gamma '$ in Bardeen-Cooper-Schrieffer theory which may be suitable
for a low-$T_{\mathrm{c}}$ superconductor. Instead of this, we use the factor $k$ as
a fitting parameter. The comparison is presented in figure~\ref{ac}.
The ac conductivity curves were fitted to equation~(\ref{conduc}) with the
normal-state conductivity measured in  \cite{Yoshi01} to be $%
\sigma _{n}=3.3\cdot 10^{6}$~$(\Omega \textrm{m})^{-1}$. The parameters we obtain
from the fit are: $H_{\mathrm{c}2}(0)=T_{\mathrm{c}}dH_{\mathrm{c}2}\left( T\right) /dT|_{T_{\mathrm{c}}}=178$~T (corresponding to $\xi =13.6$~{\AA}), the GL parameter $%
\kappa =49.7$, the order parameter effective thickness $s'=5.51$~{\AA}, and the factor $k=\sigma _{n}/\sigma _{\mathrm{GL}}=0.86$, where we
take $\gamma =10$ for YBCO in  \cite{Li02}.
Using those parameters, we obtain $\mathrm{Gi}=2.78\cdot 10^{-3}$ (corresponding to $%
\eta =0.176$). The order parameter effective thickness $s'$ can be
taken to be equal to the layer distance (see in  \cite{Poole}) of
the superconducting CuO$_{2}$ plane plus the coherence length $2\xi _{\mathrm{c}}=2%
\frac{\xi }{\gamma }$ due to the proximity effect: $3.18~\text{\AA}+2\frac{13.6}{%
10}~\text{\AA}=4.54~\text{\AA}$, roughly in agreement in magnitude with the fitting
value of $s'$.
\section{Discussion and conclusion}

The complex conductivity was calculated in a layered type-II superconductor
under magnetic field in the presence of strong thermal fluctuations on the
mesoscopic scale in linear response. While in the normal state, the
dissipation involves unpaired electrons, in the mixed phase it takes a form
of the flux flow. Time dependent Ginzburg-Landau equations with thermal
noise describing the thermal fluctuations are used to describe the
vortex-liquid regime. The nonlinear term in dynamics is treated using the
renormalized Gaussian approximation. Explicit expressions for the complex
conductivity $\sigma _{\mathrm{s}}$ and resistivity $\rho _{\mathrm{s}}$ including all Landau
levels were obtained, therefore the approach is valid for arbitrary values if the magnetic
field is not too close to $H_{\mathrm{c}1}\left( T\right) $.

The results were compared to the experimental data on HTSC. The results are in good qualitative and even quantitative agreement
with experimental data on YBCO . The thermal fluctuation was included in the present approach, so
the results should be applicable for above and below $T_{\mathrm{c}}$.
\subsection*{Acknowledgements}
We are grateful to Baruch Rosenstein, Dingping Li for discussions.
This work was supported by the National Foundation for Science and
Technology Development (NAFOSTED) of Vietnam under Grant No.~103.02--2011.15.

\clearpage

\ukrainianpart

\title{Комплексна провідність у сильно флуктуюючих шаруватих
надпровідниках}
\author{Д.Тінь, Л.М. Тхю, Л.Б. Хоа}
\address{Фізичний факультет, національний університет освіти м. Ханой, 136 Сюентхуі, Канзяі, Ханой, В'єтнам}

\makeukrtitle

\begin{abstract}
Часозалежним методом Гінзбурга-Ландау розраховано комплексну
флуктуаційну провідність у шаруватому надпровіднику II-го типу під
дією магнітного поля. Шарувата структура надпровідника враховується
за допомогою моделі Лоуренса-Доняха, тоді як член, що описує
нелінійну взаємодію в динаміці, розглядається у самоузгодженому
гаусовому наближенні. У високотемпературних матеріалах значна
частина діаграми $H-T$ належить до вихорової рідинної фази. Вирази
для сумарного внеску від усіх рівнів Ландау записані в явній формі.
Вони застосовні по суті до всієї цієї фази і порівнюються з
експериментальними даними для високотемпературного надпровідника
YBa$_{2}$Cu$_{3}$O$_{7-\delta }$. Вище кросовера до ``нормальної
фази'' наші результати узгоджуються з раніше отриманими.
\keywords часозалежний метод Гінзбурга-Ландау, комплексна
провідність, надпровідник II-го типу

\end{abstract}

\end{document}